\documentclass[preprint]{aastex}

\slugcomment{}
\shorttitle{THE MASS PROFILES AND FORMATION EPOCH OF DARK HALOS}
\shortauthors{Sato et al.}

\begin{document}

\title{THE OBSERVED MASS PROFILES OF DARK HALOS \\
    AND THE FORMATION EPOCH OF GALAXIES}

\author{Shinji Sato\altaffilmark{1}, Fumie Akimoto\altaffilmark{1}, Akihiro Furuzawa\altaffilmark{1}, Yuzuru Tawara\altaffilmark{1}, \\ and Manabu Watanabe\altaffilmark{1}}
\affil{Department of Astrophysics, Nagoya University,
    Nagoya, JAPAN}

\and

\author{Yasuki Kumai\altaffilmark{2}}
\affil{Kumamoto Gakuen University, Kumamoto, Japan}

     \begin{abstract}

    We have determined the mass profiles of dark halos in 83 objects observed by ASCA.  The point spread function of X-ray telescope was deconvoled by the Richardson-Lucy algorithm and the temperature profiles were calculated to obtain the mass profiles.  The derived mass profiles are consistent with the NFW model in $ 0.01-1.0 r_{virial}$.  We found a good correlation between the scale radius $ r_{s}$ and the characteristic mass density $ \delta_{c} $ , which indicates the self-similarity of dark halos.  The spectrum index of primordial density fluctuation, $ P(k) \propto k^{n}$, was determined from the slope of $ r_{s} - \delta_{c} $ relation.  For $M_{200}=10^{12}-10^{15} M_{\odot}$, our analysis gives $n=-1.2 \pm 0.3$ with a confidence level of $90\%$.  The mass density of dark halos is a good indicator of the mean mass density of the universe at the time when the halos were assembled, $z=z_{f}$.    Assuming $\delta _{c}  \propto (1+z_{f} )^{3}$, we have determined the epoch when each dark halo was assembled.  Our analysis indicates that the field elliptical galaxies and groups of galaxies formed approximately at $1+z_{f} \simeq 15$  and at $1+z_{f} \simeq 7-10$ respectively.   
  
\end{abstract}

\keywords{cosmology: dark matter---cosmology: observations---galaxies: clusters: general---galaxies: formation---galaxies: halos---X-ray: galaxies}

\section{Introduction}

   Early on, the density fluctuation grows linearly as the universe expands, $d\rho / \rho  \propto (1+z)^{-1}$ .  Once a density enhancement of a spherical region approaches $ d\rho /\rho \sim 1$, the sphere begins to turn around from the Hubble flow and collapse rapidly to form a virialized halo.  The over density of virialized halo reaches  $178\rho_{0}(1+z_{c})^{3}$, where $z_{c}$ is the collapse redshift and $\rho_{0}$ is the critical mass density in the present \citep{kai86}.   As the universe expands, the virial radius expands gradually by steady mass accretion.  N-body simulations show that the accreting dark matter forms an extended envelope, while the density profile of the original halo remains  unchanged.  The mass profile of a dark halo is rearranged only by a major merger through violent relaxation.  A halo formed by a major merger at $z_{m}$ is characterized by the mass density of $178\rho_{0}(1+z_{m})^{3}$, and it is preserved until the next major merger takes place \citep{sal98, rai98, hen99}.  By this means, the density profile of a dark halo can be a good indicator of the redshift when  the last merger took place.  

 \citet{nav96}  (hereafter NFW) suggested that all of the dark halos have the self-similar mass profile described as 

\begin{equation}
\rho_{(r)} = \frac{\rho_{0} \delta_{c}}{(r/r_{s}) (1+r/r_{s})^{2} },
\end{equation}

 where $\rho_{0}, r_{s}$ and $\delta_{c}$ are the critical mass density of universe, the scale radius and the characteristic density respectively. In this case,  the $\delta_{c}$ would be a direct indicator of the formation epochs of dark halos.

In this paper, we will present the mass profiles of dark halos over the mass range of  $10^{12}-10^{15} M_{\odot}$ using a sample of 83 objects observed with ASCA. This furnishes us with a possible way to determine the formation epochs of galaxies and clusters of galaxies. 

Throughout the paper, we assume $ \Omega=1$, $\Lambda =0$ and $H_{0}=75km s^{-1} Mpc^{-1}$.

\section{SAMPLE AND ANALYSIS}

 Our sample consists of rich clusters, poor clusters, groups of galaxies and elliptical galaxies.  The spectrum analysis was performed by the standard method using the $Xselect$ and the $Xspeck$ software to determine the HI absorption, the metal abundance and the luminosity weighted temperature. The results of spectrum analysis are described in the forthcoming paper along with a full description of our sample objects \citep{aki00}.

The point spread function (PSF) of the ASCA telescope is characterized by a sharp central peak (FWHM $\sim 15arcsec$) and a significant scattering tail extending to $\sim 3arcmin$.  We have deconvolved this scattering tail by the Richardson-Lucy algorithm \citep{luc74}. The observed images of 3C273 were employed as the PSFs.   The deconvolutions were performed in the three energy bands ($0.5-1.5, 1.5-3.0$ and $3.0-10keV$) separately, since the PSF has a weak dependence on the photon energy.   All of the observed images went through 100 iterations.

 We have calculated the X-ray brightness profiles from the deconvoled images, masking the bright sub-peaks and obvious sub-clusters.  The brightness profiles obtained in the lowest energy band, $ kT=0.5-1.5keV$,  were compared to those from ROSAT PSPC in the energy band of $ kT=0.5-2.0keV$, and good agreements were found in these independent observations.  

The temperature profiles are calculated from the brightness ratio of the X-ray profiles in the three energy bands, $kT = 0.5-1.5, 1.5-3.0  and  3.0-10 keV$.  We assumed that there were no radial gradients in metal abundance and HI absorption. The typical temperature profiles of our sample are shown in Figure 1, together with a comparison with the previous measurements.

\section{CALCULATION OF MASS PROFILE} 

The total mass of a dark halo can be calculated from the gas density and the temperature profiles of ICM, assuming the hydrostatic equilibrium.  

If the observed X-ray surface brightness profile $S_{(\theta)}$ is described by the $\beta-$ model, $S(\theta)=S_{0} (1+(\theta/\theta_{c})^{2} )^{-3 \beta +0.5}$,  where $\theta$, $\theta_{c}$ and $\beta$ are the angular radius, the core radius and the $\beta$ parameter respectively,  and the radial gradient of temperatures is relatively small as compared with that of the gas density, the total mass within radius r is given by

\begin{equation}
M_{(r)} = \frac{3 \beta k T_{(r)}}{G \mu m_{p}} r \frac{ ( r/r_{c} )^{2} } {1+(r/r_{c} )^{2}  } ,
\end{equation}

where $ T$, $k$, $\mu m_{p}$  and $r_{c}$ are the gas temperature, the Boltzmann$^\prime$s constant, the mean molecular weight of the hot gas and the core radius respectively.

We found that $r_{c}$, $\beta$ and $kT$ were not constant in a single cluster, but changed slowly with radius.  The angular profiles of two parameters, $r_{c (\theta)}$ and $\beta_{(\theta)}$, were calculated  by fitting the $\beta-$ model to the local brightness profiles of $\theta\pm 0.3 \theta$.  Although the angular profiles, $r_{c (\theta)}$, $\beta_{(\theta)}$ and $kT_{(\theta)}$, are the luminosity weighted  properties of the hot gas along the line of sight, these are nearly equal to the radial profiles, $r_{c (r)}$, $\beta_{(r)} $ and $kT_{(r)}$, if the radial gradient of gas density (or $S_{(\theta)}$) is much larger than those of $r_{c}$ , $\beta$ and $kT$.  In our sample, $r_{c (\theta)}$, $\beta_{(\theta)}$ and $kT_{(\theta)}$ change factor 2 at most, while the $S_{(\theta)}$ changes 2 or 3 orders of magnitude. We therefore assumed  that the observed angular profiles represented the radial profiles of these parameters. The local gas density profile was therefore given by $n_{(r)} = n_{0} (1+(r/r_{c (r)})^{2})^{-3\beta_{(r)} /2} $ and the total mass $M_{(r)}$ was calculated from the equation (2). The mass density $\rho_{(r+0.5dr)} = (M_{(r+dr)}-M_{(r)} )/(4 \pi r^{2} dr)$ was obtained from $M_{(r)}$ and $M_{(r+dr)}$, keeping $ dr=26arcsec$.

 We have constructed the models of gas halos bound by the NFW dark halos to evaluate the systematic errors in our method. The mass profiles were calculated from the model profiles, which correspond to typical galaxies, poor clusters and rich clusters, by applying the same method. The best-fit $\delta_{c}$ and $r_{s}$ are then calculated by fitting the NFW model to the mass profiles. Comparing the calculated  $\delta_{c}$ and $r_{s}$ with the original values, we have confirmed that the systematic errors of our method were much smaller than the typical photon noise of our sample objects.

 In our analysis, the brightness profile within $\theta<0.5 arcmin$ were excluded to avoid the contribution of the cooling flow component.   The maximum radii were extending up to $10-40 arcmin$ depending on the photon number available. We define $r_{200}$ as the radius where the mean interior density becomes 200 times the critical mass density of the universe, and $M_{200}$ as the total mass within  $r_{200}$.  In this paper, we consider $r_{200}$ and $M_{200}$ is the virial radius and the virial mass \citep{col96} respectively.

\section{THE OBSERVED MASS PROFILES} 

 The NFW model was fitted to our mass profiles to determine $\delta_{c}$ , $r_{s}$,  $r_{200}$ and $M_{200}$ .  The mass profiles of several objects, which have similar characteristic radius,  are co-added to improve the signal to noise ratio.  The composite mass profiles, scaled by $r_{s}$ are shown in Figure 2 by the shaded lines.   The widths of lines correspond to the $\pm 1\sigma$ errors.  The average scale radii are $<r_{s}>=24, 59, 150, 200$ and  $460kpc$ in (a), (b), (c) and (d) respectively.  The NFW model and the $\beta -$ model were fitted to these composite profiles.  Since the composite mass profiles drop more quickly at larger radii,  the NFW model  gives better fit to the observations.  The $\chi^{2}/d.o.f.$ of the NFW model are  3.0/4, 4.3/5, 5.5/5 and 3.6/5 in (a), (b), (c) and (d), while those of the $\beta -$ model are 2.0/4, 24/5, 23/5 and 14/5.  To illustrate a similarity of  mass profile, we have normalized the composite density profiles by $ \rho_{(r_{s})} / 10^{9} $ and illustrated in the upper portion of Figure2 by the solid lines.   It is remarkable that these haloes have very similar mass profiles.  

The brightness profiles scaled by $kT^{1/2} (1+z)^{9/2}$  can be a good indicator of mass profiles.   \citet{pon99} illustrated the scaled brightness profiles of 25 clusters, normalized by the viral radius, and pointed out the systematic change of profiles with $kT^{1/2} (1+z)^{9/2}$.    Our measurements, however, do not conflict with this result.  Since the observed dark halos have different concentration parameters $c=r_{200}/r_{s}$  with $M_{200}$ as shown in the upper panel of Figure 3, the virial radius is not an adequate scale to illustrate the similarity of mass profiles.  If  our results are scaled by $r_{200}$, we see the same systematic trend as found by \citet{pon99}.

The correlation between the $M_{200}$ and the luminosity weighted temperature $kT$ is shown in the lower panel of Figure 3.  Our measurements give $kT \propto M_{200}^{0.49 \pm 0.10} $ with a confidence level of $90 \%$. 
We have found a good correlation between $\delta_{c}$ and $r_{s}$ in our 83 objects as shown in Figure 4.   Our result is essentially the same as the $c-M_{vir}$ and $\delta_{c}-M_{vir}$ correlations found by \citet{wuu00},  but have larger dynamic ranges and smaller errors.  The upper panel indicates residuals from the best-fit model.  The rms scattering of $\delta_{c}$ shown by the dashed lines ( $ \pm 45 \%$ ) is significantly larger than the $ 1 \sigma $ errors of measurements.  Our results indicate that the mass profile of a dark halo is described by a single parameter, such as $r_{s}, \delta_{c}$ or $ M_{200}$.

\section{THE SPECTRUM INDEX OF PRIMORDIAL FLUCTUATION AND THE FORMATION EPOCH OF GALAXIES} 

The amplitude of primordial density fluctuation, $P(k) \propto k^{n}$, is characterized by the spectrum index $n$.  In the CDM universe, the radius and the mass density of dark halo are connected to $n$ as $\rho \propto r^{-(3n+9)/(n+5)} $ \citep{pee80}.  The slope of $ r_{s} - \delta_{c}$ relation shown in Figure 4 is therefore the direct indicator of the power spectrum.  If the $P(k)$ is described by a single power in the mass range of  our sample $M_{200}=10^{12}-10^{15} M_{\odot}$, the $ r_{s} - \delta_{c}$ relation gives $n=-1.2 \pm 0.3$ with a confidence level of $90 \%$.  The error is calculated from the intrinsic scattering of the objects.  When the objects of $M_{200} <2 \times 10^{14} M_{\odot}$ are excluded, we get  $n=-0.8 \pm 0.7 (90 \%)$, which is consistent with the result of \citet{wuu00}, $n=-0.7 \pm 0.3 (1\sigma)$.

We assume that the characteristic density $ \delta_{c}$ is preserved during the steady accretion phase as N-body simulations suggest.  In that case,  $ \delta_{c}$ should be proportional to the mean mass density of the universe when the halo was assembled, $\delta_{c}= f (1+z_{f})^{3 }= f (1+z_{obj})^3  [ (1+z_{f})/(1+z_{obj})]^3  $, where $f$, $1+z_{f}$ and $1+z_{obj}$ are a proportional constant, the formation epoch and the observed redshift of object respectively.  As $1+z_{f} \geq 1+z_{obj}$, $\delta_{c}/(1+z_{obj})^{3}= f [(1+z_{f})/(1+z_{obj})]^{3}  \geq f $.   If our sample includes the zero age clusters $z_{f} = z_{obj}$,  the $\delta_{c}/(1+z_{obj})^{3}$ of these objects would be equal to $f$.  As shown in Figure 5, we found the clear locus in the lower end of  $ (1+z_{obj}) $ v.s.  $\delta_{c}/(1+z_{obj})^{3}$  plot, and five clusters on the locus (A370, A1758, CL0016+16, MS04516 and RXJ1347.5-1145) were designated to the zero age clusters.   These zero age clusters give the proportional constant and the formation epoch as $f=1640 \pm 260 $ and $ 1+z_{f} = (0.085  \pm 0.005) \delta_{c}^{1/3} $.  Applying this equation to individual objects, we get $1+z_{f} =13-17 $ for the formation epochs of 3 elliptical galaxies in our sample (NGC1399, NGC3923 and NGC4636).    Our results are nearly independent of the cosmological parameters as illustrated in \citet{wuu00}.

\acknowledgments

We are grateful to the ASCA team for their efforts on the design and operation of ASCA hardware and software.

\clearpage
\begin{figure}
\figurenum{1}
\epsscale{0.75}
\plotone{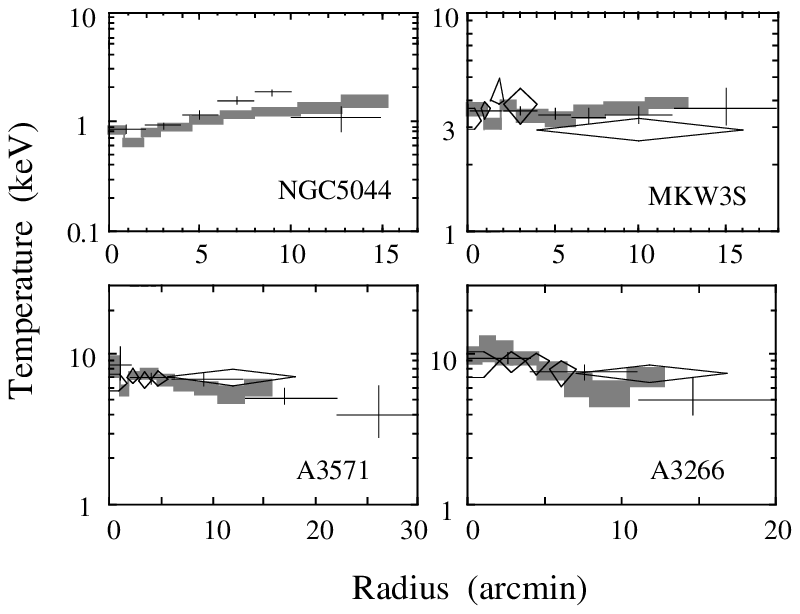}
\caption{The measured temperature profiles of ICM.  The shaded boxes are our results.   The height of the boxes indicates the error of $90\%$ confidence level.  The crosses are the results from \citet{fuk96} (NGC5044), \citet{kik99} (MKW3S) and \citet{mar98} (A3571 and A3266).  The vertical lines correspond to the $90\%$ errors.  The diamonds are from  \citet{whi00}.  The heights of diamonds indicate the $ 1 \sigma $ errors.  }
\end{figure}

\clearpage
\begin{figure}
\figurenum{2}
\epsscale{0.7}
\plotone{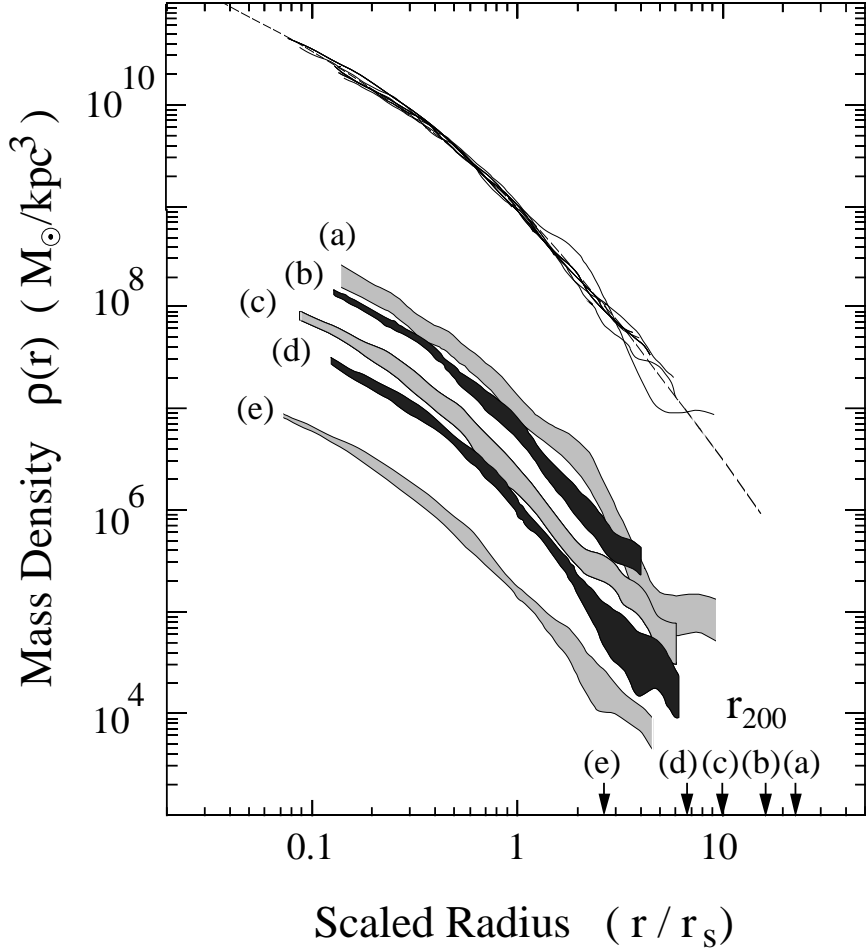}
\caption{The composite mass profiles scaled by $r_{s}$.  The width of shaded lines indicates the $ 1 \sigma $ errors.    The average scaling factors of emission measure are $kT^{1/2} (1+z)^{9/2}=0.88, 2.3, 3.6, 6.1$ and $9.0 $ in (a), (b), (c), (d) and (e) respectively.  The solid lines are mass profiles scaled by $r_{s}$ and $ \rho_{(r_{s})}/ 10^{9}$.  The dashed line is the best-fit NFW model.  The arrows indicate the virial radii $r_{200}$.}
\end{figure}

\clearpage
\begin{figure}
\figurenum{3}
\epsscale{0.7}
\plotone{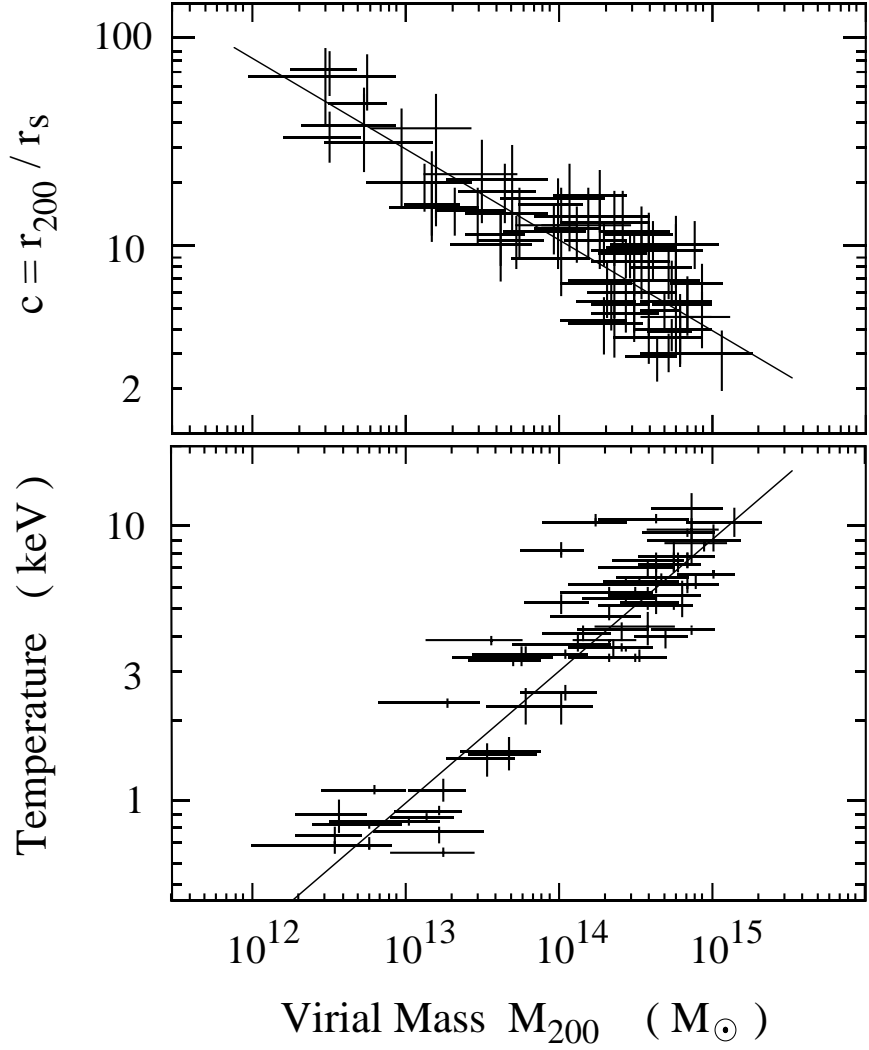}
\caption{The upper and lower panels are  the measured $M_{200} - c$ and $ M_{200} - kT$ relations. The $kT$ is the luminosity weighted temperature determined by the X-ray spectrum. The error bars correspond to the $90\%$ errors.  The best fit relations with $90\%$ error are $log c = (7.2 \pm 1.8) - (0.44 \pm 0.13) log M_{200}$ and $log kT = (-6.4 \pm 1.4) + (0.49 \pm 0.10) log M_{200}$ .}
\end{figure}

\clearpage
\begin{figure}
\figurenum{4}
\epsscale{0.7}
\plotone{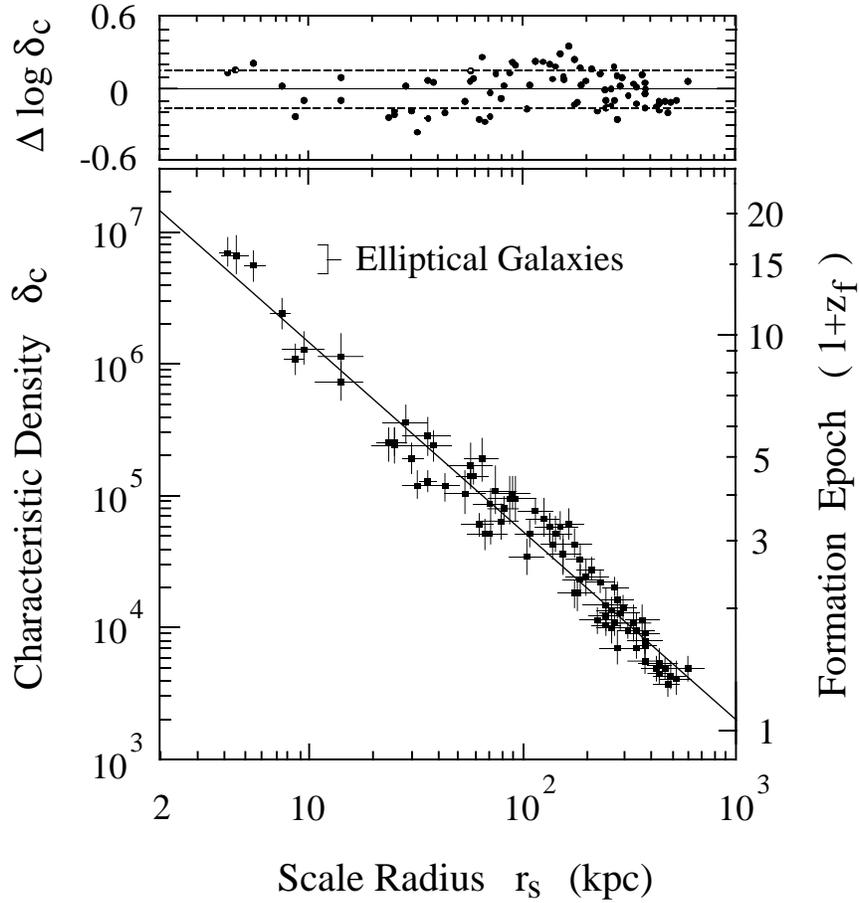}
\caption{The measured $\delta_{c} - r_{s}$ relation and the best-fit model.  The error bars correspond to the $90\%$ errors.  The residuals are shown in the upper panel.  The formation epoch of dark halos is indicated in the right side. The best fit relation with $90\%$ error is $log \delta_{c} = (7.61 \pm 0.29) - (1.44 \pm 0.14) log r_{s} $.}
\end{figure}

\clearpage
\begin{figure}
\figurenum{5}
\epsscale{0.7}
\plotone{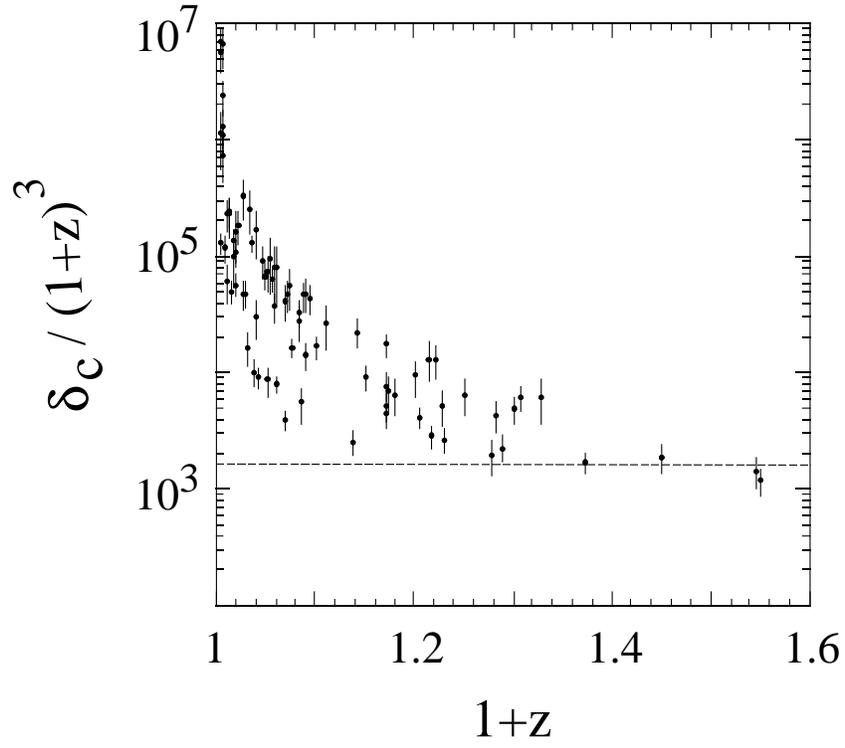}
\caption{A finding method for the zero age clusters. The $(1+z)$ is the redshift of each object and the $\delta_{c} / (1+z)^{3} $ indicate the age of dark halo since the last assembly. The dashed line is the average of 5 clusters on the lower boundary of  $\delta_{c} / (1+z)^{3} - (1+z)$  distribution.}
\end{figure}


\begin{thebibliography}{}


\bibitem[Akimoto et al.(2000)]{aki00} Akimoto, F., Furuzawa, A., Kumai, Y., Sato, S., Tawara, Y., \& Watanabe, M.  2000,  in preparation.

\bibitem[Carlberg et al.(1997)]{car97} Carlberg, R.  G., et al.  1997, \apj, 485, L13

\bibitem[Cole \& Lacey(1996)]{col96} Cole, S., \&
    Lacey, C.   1996, \mnras, 281, 716


\bibitem[Fukazawa et al.(1996)]{fuk96} Fukazawa, Y., et al.   1977, \pasj, 48, 395
  
\bibitem[Henriksen \& Widrow(1999)]{hen99} Henriksen, R.  N., \& Widrow, L.  M.  1999, \mnras, 302, 321

\bibitem[Kaiser(1986)]{kai86} Kaiser, N.  1986, \mnras, 222, 323

\bibitem[Kikuchi et al.(1999)]{kik99} Kikuchi, K., Furushyo, T., Ezawa, H., Yamasaki, N.  \&  Ohashi, T.   1999, \pasj, 51, 301

\bibitem[Lucy(1974)]{luc74} Lucy, L.  B.  1974, \apj, 79, 745 

\bibitem[Markevitch(1998)]{mar98} Markevitch, M., Forman, W.  R.  , Sarazin, C.  L.  \& Vikhlinin, A.  1998, \apj, 503, 77

\bibitem[Navarro, Frenk \& White(1996)]{nav96} Navarro, J.  F., Frenk, C.  S., \& White, S.  D.  M.  1996, \apj, 462, 563

\bibitem[Peebles(1980)]{pee80} Peebles, P.  J.  E.  The Large-Scale Structure of the Universe, Princeton:Princeton Univ.  Press, 1980

\bibitem[Ponman, Cannon \& Navarro(1999)]{pon99} Ponman, T.  J., Cannon, D.  B., \& Navarro, J.  F.  1999, \nat, 397, 135

\bibitem[Raig, Gonz\'alez-Casado \& Salvador-Sol\'e(1998)]{rai98} Raig, A., Gonz\'alez-Casado, G.,\& Salvador-Sol\'e, E.  1998, \apj, 508, L129

\bibitem[Salvador-Sol\'e, Solanes \& Manrique(1998)]{sal98} Salvador-Sol\'e, E., Solanes, J.  M., \& Manrique, A.  1998, \apj, 499, 542

\bibitem[White(2000)]{whi00} White, D.  A.   2000, \mnras, 312, 663

\bibitem[Wu \& Xue(2000)]{wuu00} Wu, X.  -P., \& Xue, Y.  -J.   2000, \apj, 529, L5


\end{thebibliography}
\end{document}